\documentclass[sn-mathphys,Numbered]{sn-jnl}

\usepackage{graphicx}%
\usepackage{multirow}%
\usepackage{amsmath,amssymb,amsfonts}%
\usepackage{amsthm}%
\usepackage{mathrsfs}%
\usepackage[title]{appendix}%
\usepackage{xcolor}%
\usepackage{textcomp}%
\usepackage{manyfoot}%
\usepackage{booktabs}%
\usepackage{algorithm}%
\usepackage{algorithmicx}%
\usepackage{algpseudocode}%
\usepackage{listings}%
\usepackage{CJKutf8}

\theoremstyle{thmstyleone}%
\theoremstyle{thmstyletwo}%

\theoremstyle{thmstylethree}%

\begin{document}
\begin{CJK}{UTF8}{gbsn}

\title[Article Title]{A Deep Learning-Driven Pipeline for Differentiating Hypertrophic Cardiomyopathy from Cardiac Amyloidosis Using 2D Multi-View Echocardiography}


\author[1,2,3]{\fnm{Bo} \sur{Peng}}\email{bopeng@swpu.edu.cn}
\author[1]{\fnm{Xiaofeng} \sur{Li}}
\author[1]{\fnm{Xinyu} \sur{Li}}
\author[1]{\fnm{Zhenghan} \sur{Wang}}
\author[1]{\fnm{Hui} \sur{Deng}}
\author[2,4]{\fnm{Xiaoxian} \sur{Luo}}
\author[2,3]{\fnm{Lixue} \sur{Yin}}
\author*[2,3]{\fnm{Hongmei} \sur{Zhang}}\email{oiczhm@163.com}

\affil[1]{\orgdiv{School of Computer Science and Software Engineering}, \orgname{Southwest Petroleum University}, \orgaddress{\street{No. 8 Xin Du Avenue}, \city{Chengdu}, \postcode{610500}, \country{People’s Republic of China}}}

\affil[2]{\orgdiv{Department of Cardiovascular Ultrasound and Noninvasive Cardiology}, \orgname{Sichuan Provincial People's Hospital, University of Electronic Science and Technology of China}, \orgaddress{\street{No. 32 West Second Ring Road}, \city{Chengdu}, \postcode{610072}, \country{People’s Republic of China}}}

\affil[3]{\orgdiv{Ultrasound in Cardiac Electrophysiology and Biomechanics Key Laboratory of Sichuan Province}, \orgname{Sichuan Provincial People's Hospital, University of Electronic Science and Technology of China}, \orgaddress{\street{32 West Second Ring Road}, \city{Chengdu}, \postcode{610072}, \country{People’s Republic of China}}}

\affil[4]{\orgdiv{School of Medical and Life Sciences}, \orgname{Chengdu University of Traditional Chinese Medicine}, \orgaddress{\street{No. 37 Twelve Bridge Road}, \city{Chengdu}, \postcode{610075}, \country{People’s Republic of China}}}

\abstract{Hypertrophic cardiomyopathy (HCM) and cardiac amyloidosis (CA) are both heart conditions that can progress to heart failure if untreated. They exhibit similar echocardiographic characteristics, often leading to diagnostic challenges. This paper introduces a novel multi-view deep learning approach that utilizes 2D echocardiography for differentiating between HCM and CA. The method begins by classifying 2D echocardiography data into five distinct echocardiographic views: apical 4-chamber, parasternal long axis of left ventricle, parasternal short axis at levels of the mitral valve, papillary muscle, and apex. It then extracts features of each view separately and combines five features for disease classification. A total of 212 patients diagnosed with HCM, and 30 patients diagnosed with CA, along with 200 individuals with normal cardiac function(Normal), were enrolled in this study from 2018 to 2022. This approach achieved a precision, recall of 0.905, and micro-F1 score of 0.904, demonstrating its effectiveness in accurately identifying HCM and CA using a multi-view analysis.}




\maketitle

\section{Introduction}\label{sec1}

Cardiomyopathy is defined by structural and functional abnormalities in the ventricular myocardium. This condition may lead to heart failure, arrhythmias, syncope, chest pain, and sudden cardiac death (SCD). Hypertrophic Cardiomyopathy (HCM) stands out as one of the most prevalent forms of hereditary primary cardiomyopathy. Statistics indicate that the crude prevalence of HCM in the Chinese population is approximately 0.16, with an adjusted prevalence of 80 per 100,000 after considering age and gender factors \cite{bib2}. Consequently, it is estimated that there are over one million adult HCM patients in China \cite{bib2}. Cardiac amyloidosis (CA) is the most common type of secondary restrictive cardiomyopathy. Amyloidosis is a group of diseases in which insoluble amyloid deposits in organs, leading to dysfunction of organ and tissue. The heart is an organ often affected by amyloidosis, usually manifested as myocardial hypertrophy and restrictive congestive heart failure \cite{bib3}.

Echocardiography stands as a highly effective and commonly used diagnostic tool for a range of heart diseases, notably in the screening of HCM and CA \cite{bib5}. In generally, patients with HCM typically exhibit marked hypertrophy of the myocardium, most notably of the interventricular septum, leading to a thickened heart wall, particularly in the left ventricle. It often results in reduced chamber size and impaired diastolic function. On echocardiography, HCM typically presents as enhanced myocardial texture, with wall thickness generally exceeding 15mm and often distributed unevenly \cite{bib34}. While HCM is primarily marked by significant myocardial thickening, CA often displays a slight different echocardiographic pattern. Patients with CA typically exhibit a mild increase in wall thickness, usually less than 15mm with even distribution, mostly less than 15mm and evenly distributed. Occasionally, the interventricular septum and right ventricle may be affected. Consequently, CA is often characterized by diffuse, punctate echoes slightly distributed within the myocardium on echocardiography \cite{bib33}.It should be noted that while cases of HCM with pericardial effusion are relatively rare, cases of CA with slight pericardial effusion are relatively more common.

Despite these distinctions, the two conditions are often challenging to differentiate echocardiographically due to their similar presentation of myocardial thickening. The overlap in their imaging features, especially in less pronounced cases, can lead to misdiagnosis \cite{bib4}. Additionally, the rarity of CA compared to HCM contributes to its less frequent consideration in differential diagnoses. This similarity in echocardiographic findings, combined with the subtleties in visual cues and the necessity for a highly skilled interpretation, underscores the diagnostic challenge in distinguishing between these two cardiac conditions. 

Given the similarities in echocardiographic findings between HCM and CA, accurate diagnosis heavily relies on the operator's expertise, necessitating substantial professional knowledge due to the complex operation and intricate measurements involved. Expert sonographers, through a combination of techniques such as tissue Doppler, mitral valve spectrum analysis, and assessment of clinical symptoms, can often differentiate between these conditions. However, in cases of uncertainty, myocardial biopsy may be required to confirm the diagnosis. This complexity leads to an extended training period for cardiac ultrasound physicians and notable variability in diagnostic proficiency among practitioners \cite{bib6}. Research indicates that the misdiagnosis rate for cardiovascular diseases in medical imaging ranges from 10\% to 30\%, with diagnostic sensitivity varying between 27.5\% and 96\% across different medical institutions \cite{bib7,bib8}. 


Recent advancements in deep learning-based artificial intelligence (AI) have significantly improved medical image analysis, particularly in cardiovascular diagnostics. This progress has led to key developments in AI-assisted echocardiographic analysis, playing a crucial role in enhancing 'AI in Cardiovascular Diagnostics'. Significant achievements include advancements in diagnosing cardiomyopathy \cite{bib10,bib11,bib12,bib13}, identifying severe coronary stenosis \cite{bib14}, evaluating the severity of mitral regurgitation, detecting the presence of prosthetic valves \cite{bib15,bib16}, analyzing left ventricular wall thickness \cite{bib19}，and enhancing overall heart disease detection \cite{bib17,bib18}. Specifically, in the differentiation between HCM and CA, leveraging the ability to distinguish differences in myocardial echogenicity, wall thickness and distribution, and pericardial conditions through the apical 4-chamber view and the parasternal long axis of the left ventricle, Yu et al. \cite{bib0} innovatively developed a three-tier classification network to discern three types of left ventricular hypertrophy - hypertensive heart disease, HCM, and CA by integrating features from the two views. Despite the innovation, the reported accuracy of this network stands at 75\%. This indicates a significant scope for improvement to meet the stringent accuracy standards required for clinical cardiac disease diagnosis. More recently, Wu et al. \cite{bib20} introduced a multi-modality approach, achieving a remarkable maximum accuracy of 0.98 in distinguishing HCM from CA. This approach integrates clinical characteristics, conventional echocardiography, and two-dimensional speckle tracking echocardiography (2D-STE). The fundamental aspect of this approach lies in the calculation of strain and strain rate from 2D-STE used to detect myocardial deformation, which are found to be more sensitive indicators compared to conventional echocardiography. This sensitivity enables effective differentiation and treatment of HCM and CA at an early stage \cite{bib31,bib32}. It is important to note that the computation of strain and strain rate, which involves the gradient of estimated displacements from STE, is typically computationally intensive and often conducted offline. Moreover, it demands high standards in data quality, resolution, frame rate, as well as operator proficiency. These factors are crucial for ensuring the accuracy and effectiveness of the analysis \cite{bib26}.

To improve performance and address the limitations of existing methods, we introduce an automated analytical pipeline for the accurate diagnosis of HCM and CA. Our proposed pipeline initiates with the application of a Vision Transformer (ViT) to capture five distinct echocardiographic views. Subsequently, features from these views are extracted using a modified ResNet neural network. Finally, these extracted features are combined and fed into a linear network classifier, forming a three-type classification model. To our knowledge, this unique pipeline and the application of a multi-view technique for diagnosing HCM and CA have not yet been reported in the literature.

\section{Methods}\label{sec3}

\subsection{Study population}\label{subsec0}
Echocardiography images of patients were collected from the Sichuan Provincial People's Hospital, Affiliated Hospital of the University of Electronic Science and Technology of China between 2018 and 2022. This study was approved by the Ethics Committee of the Sichuan Provincial People's Hospital.

\begin{figure*}
    \centering
    \includegraphics[width=0.5\textwidth]{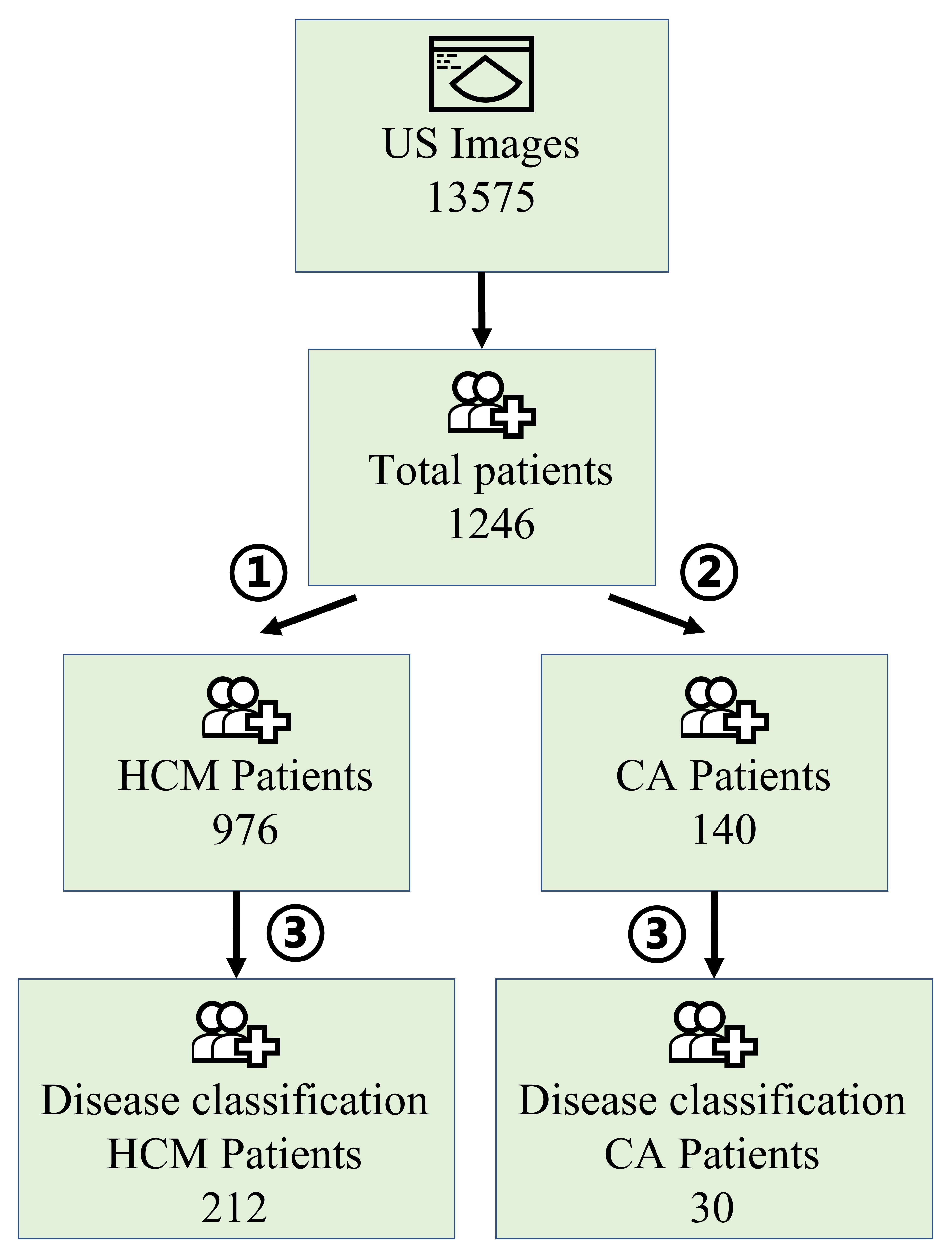}
    \caption{The schematic diagram for criteria of inclusion and exclusion criteria patients. The number of patients used for each step is indicated. ①②③ is the criterion. ① Left ventricular wall thickness ≥15mm in any segment or ≥12mm in patients with family history. ② Suspected CA by echocardiography, confirmed amyloidosis by tissue biopsy, or confirmed by late gadolinium enhanced cardiac magnetic resonance (CMR) imaging. ③ With all 5 views.
}
    \label{fig:1}
\end{figure*}

Experienced echo doctors and cardiologists (each with more than 10 years of clinical experience) utilized examination reports and corresponding clinical data to determine the diagnosis and etiology of HCM and CA. Excluding hypertension and athlete's heart, HCM was defined by linear method as left ventricular wall thickness ≥15mm in any segment or ≥12mm in patients with family history. Diagnostic criteria for CA: suspected CA by echocardiography, confirmed amyloidosis by tissue biopsy, or confirmed by late gadolinium enhanced cardiac magnetic resonance (CMR) imaging. A total of 1246 patients were included in the study. The schematic diagram for the specific screening and inclusion of studies is shown in Fig. \ref{fig:1}, which provides the number of patients in each step.

\subsection{Echocardiography}\label{subsec1}
\begin{figure*}
    \centering
    \includegraphics[width=0.5\textwidth]{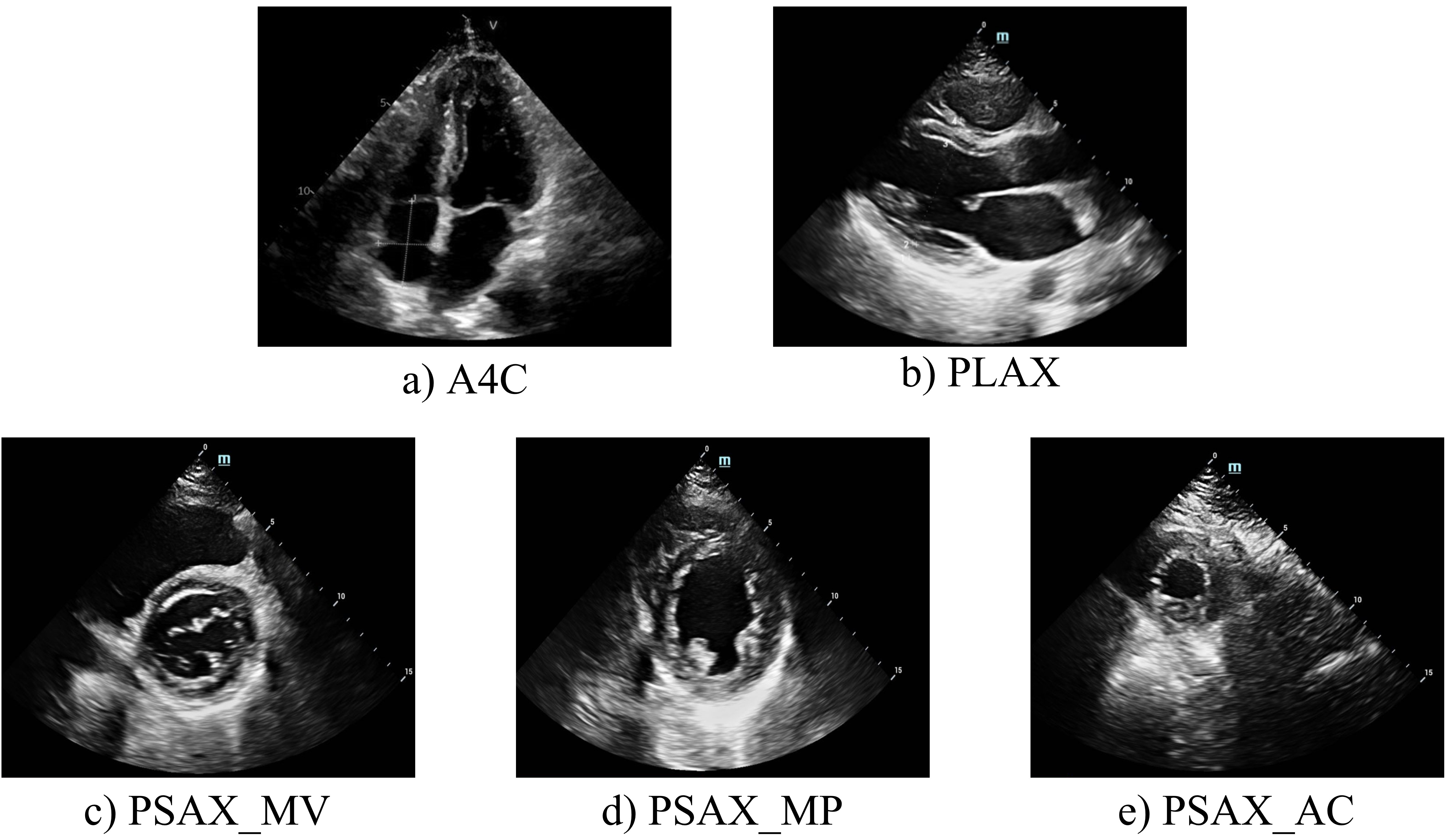}
    \caption{Cardiac imaging of five views. The second row is the short axis of arasternal left ventricular. a) A4C: apical 4-chamber, b) PLAX: Long axis of parasternal left ventricula, c) PSAX\_MV: short axis at mitral valve, d) PSAX\_MP: short axis at papillary muscle, e) PSAX\_AC: short axis at apical}
    \label{fig:2}
\end{figure*}

For each patient, five standard echocardiographic views were acquired (Fig. \ref{fig:2}), with a set of normal cardiac function(Normal) data serving as the control group. These two-dimensional echocardiograms were captured by skilled echo doctors using various equipment and were stored in JPG format.

Within these five echocardiographic perspectives, the apical 4-chamber (A4C) view (as shown in Fig. \ref{fig:2} a) is particularly effective for evaluating the dimensions of each cardiac chamber and measuring the thickness of the left ventricular wall. The long-axis view of the parasternal left ventricle (PLAX) (Fig. \ref{fig:2} b) is instrumental in examining the left ventricle's septum and posterior wall, among other structural aspects. This view plays a significant role in detecting various cardiomyopathy forms. The short axis series view of the heart (Fig. \ref{fig:2} c, d, e) provides a detailed perspective on the left ventricle's wall thickness, echogenicity, motion amplitude, and segmental thickening rate. This comprehensive insight aids in the thorough understanding of the ventricular wall and enhances the identification of specific cardiomyopathy types. By functionally evaluating these views, we aim to distinguish and diagnose HCM and CA by integrating the characteristics observed across the selected five views.

\subsection{Deep learning models}\label{subsec2}
\begin{figure*}
    \centering
    \includegraphics[width=1\textwidth]{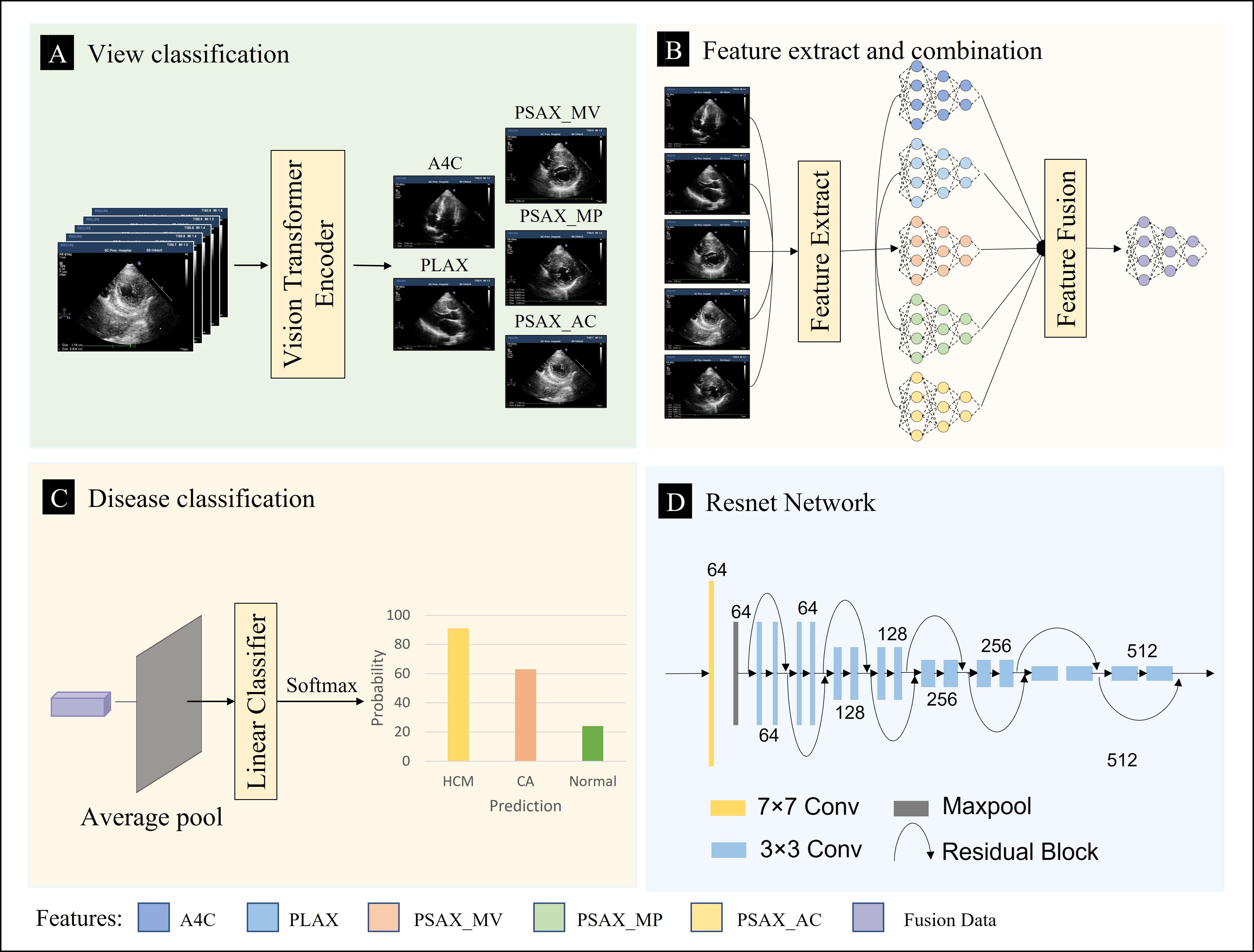}
    \caption{The pipeline for fully automated disease classification. The view classification model (A), the feature extraction model (B), three type disease classifier (C) and the network structure of feature extraction (D).}
    \label{fig:3}
\end{figure*}

The fundamental objective of this research is to establish an automated analytic pipeline for categorizing HCM and CA, serving as a supplementary diagnostic tool that operates autonomously, without the need for manual intervention. The pipeline has three stages, as illustrated in Fig. \ref{fig:3}.  The first stage used a Vision Transformer (ViT) as a classifier to classify five views: A4C, PLAX, PSAX\_MV, PSAX\_MP, PSAX\_AC of same patient (Fig. \ref{fig:3} A). The second stage then applied convolutional Resnet to extract the features of five views respectively, and fuse five features (Fig. \ref{fig:3} B). The third stag finally achieved the classification of HCM and CA through linear classifier which use the feature fused as input (Fig. \ref{fig:3} C).

\subsubsection{View classification of echocardiogram}\label{subsubsec1}
In our pipeline, the Vision Transformer (ViT) was chosen for classifying echocardiographic views. Originating from the Transformer architecture, commonly applied in natural language processing (NLP) for sequence processing \cite{bib23}, Dosovitskiy A et al. innovatively extended the Transformer model for image processing, leading to the development of ViT \cite{bib24}. The ViT approach involves segmenting images into multiple patches and creating linear embeddings from these segments. These embeddings are then processed by the Transformer as input sequences, akin to sequence processing in NLP.

\begin{figure*}
    \centering
    \includegraphics[width=0.5\textwidth]{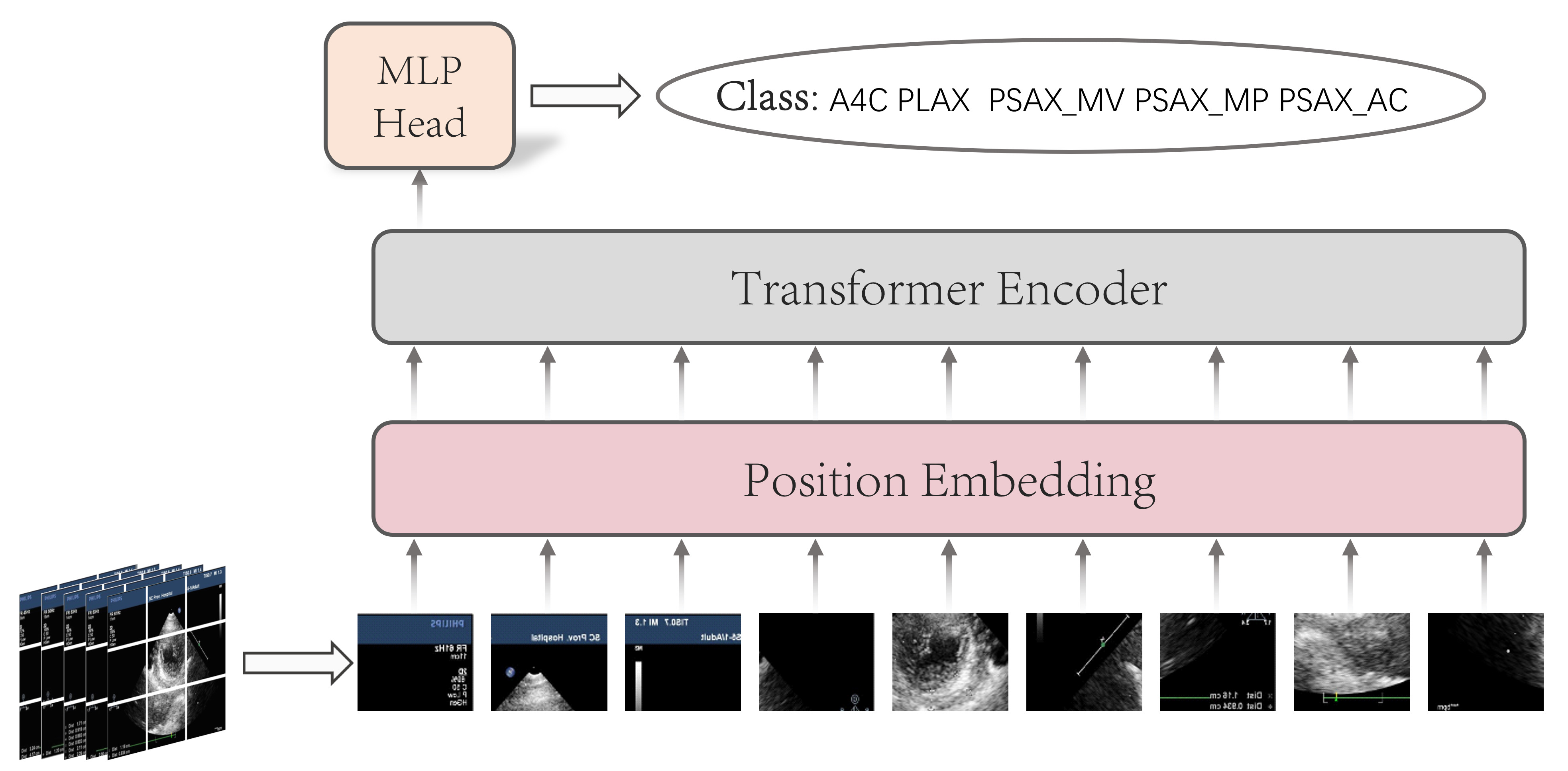}
    \caption{Transformer architecture for echocardiogram view classification. The figure illustrates how each echocardiogram is divided into several patches. These patches are then converted into linear embeddings, effectively simulating the sequence-based input that is a standard approach in NLP.}
    \label{fig:4}
\end{figure*}

The workflow of the view classification network is shown in Fig. \ref{fig:4}. In adapting the model for echocardiogram, a pivotal modification involved altering the input channels of the initial convolutional step from three to one, to better suit the grayscale format of these images. The procedure commences by dividing a 224x224 ultrasound image into 16x16 patches. Each patch is then repeated into a 16x16x3 vector. Positional encoding vectors are subsequently added to these vectors to create a comprehensive vector representation for each patch. These vector representations are then fed into the Transformer Encoder for advanced processing. Finally, a Multi-Layer Perceptron (MLP) classification head is applied to generate confidence scores for the respective classes.
\subsubsection{Multi-view feature extraction and fusion}\label{subsubsec2}
In the second stage, we chose ResNet, a residual neural network introduced in 2015 \cite{bib25}, for the purpose of feature extraction. The core principle of ResNet lies in its use of residual blocks. These blocks incorporate shortcuts that enable bypassing one or more layers of non-linear transformations. This design allows for lossless information propagation across the network, facilitating easier learning of identity mappings by directly transmitting the input signal to subsequent layers. Although both ResNet and ViT utilize skip connections, ViT tends to exhibit more overfitting than ResNet when trained on smaller datasets\cite{bib24}. Considering this characteristic, ViT is selected for the view classification stage, where it excels in capturing global contextual information from multiple views. On the other hand, ResNet is chosen for the feature fusion stage, leveraging its robustness and effectiveness in handling feature representations across different views.

In our setup, ResNet18 was chosen as the primary architecture for feature extraction. We modified it by removing the pooling and fully connected layers at the end, enabling extraction of hidden representations from input images (as shown in Fig. \ref{fig:3} D). The model first resizes all input images to 224×224 and adapts single-channel ultrasound images into a three-channel format. Using this customized ResNet18 structure, a single feature tensor of dimensions H$\times$W$\times$C = 7$\times$7$\times$512 is obtained.

Finally, the five views were fed into the ResNet18 model to extract their respective features. These features were then merged to construct a feature tensor, dimensioned at H$\times$W$\times$C = 7$\times$7$\times$2560. This tensor serves as the fused feature set for our classifier.
\subsubsection{Disease classification}\label{subsubsec3}

Disease classification is achieved through the training of a linear classifier. In this phase, the classifier processes a combined feature vector, utilizing adaptive average pooling and dimensionality reduction techniques. This model functions as a three-type classification model, effectively differentiating between HCM, CA, and Normal.

Adaptive average pooling, a technique employed to resize input feature maps to a pre-determined output dimension, offers a significant advantage over traditional fixed-size pooling methods. It dynamically adjusts the pooling operation based on the dimensions of the input, enabling neural networks to accommodate varying input sizes and produce consistent output dimensions. This approach not only increases computational efficiency but also improves the generalization ability of the model. 

Following the pooling operation, the feature tensor is reshaped to H$\times$W$\times$C = 1$\times$1$\times$2560. This is then followed by a dimensionality reduction of the pooled features, transforming them into a vector of 2560 dimensions. This vector forms the basis for training our disease classifier.

In the final step, this processed vector is utilized as the input for the linear classifier. The linear classifier operates through a process of linear mapping, converting input features into distinct class scores. Owing to their simplicity, ease of implementation, and minimal computational and storage demands, linear classifiers are particularly effective in few-shot classification learning scenarios. Through the linear classifier, the HCM, CA and Normal were classified.

\section{Experiments}\label{sec4}
\subsection{Dataset}\label{subsec1}

In the view classification phase, we utilized 13,575 ultrasound echocardiographic images from 1246 patients. These were carefully categorized by experienced echo doctors into various views and were thoroughly reviewed. The distribution was as follows: 3,980 images in the A4C view, 4,500 in PLAX, 1,417 in PLAX\_MV, 1,815 in PLAX\_MP, 1,863 in PLAX\_AC, and 2,686 images classified as other views. For the purposes of training and evaluation, the dataset was divided into training, validation, and test sets following an 8:1:1 ratio. Table \ref{tab1} details the data distribution for each specific view.
\begin{table*}[h]
\caption{Number of the private dataset used for view classification training and validation datasets}\label{tab1}%
\begin{tabular}{@{}lllllll@{}}
\toprule
Datasets       & A4C  & PLAX  & PSAX\_MV  & PSAX\_MP & PSAX\_AC & Other\\
\midrule
Training set   & 3184 & 3600  & 1133     & 1452    & 1490    & 2686 \\
Validation set & 398  & 450   & 143      & 182     & 187     & 270 \\
Test set       & 398  & 450   & 141      & 181     & 186     & 268 \\
\botrule
\end{tabular}
\end{table*}

In the feature extraction and disease classification phase, we selected patients with five distinct ultrasound views from 1246 patients, a total of 242 patients for both training and validation. These cases were divided into two major disease categories: 212 cases of HCM and 30 cases of CA. For model training and validation, the dataset was divided according to an 8:2 ratio for training and validation, respectively. Subsequently, specific augmentation methods, including rotating 15 degrees, gaussian blur, contrast enhancement, brightening adjustment and dimming, were applied to the training and validation sets of CA independently. Following this process, the training set of CA was expanded to 144 samples, and the validation set to 36 samples. These augmentations are detailed in Table \ref{tab2}. This approach was chosen to maximize the model's effectiveness and its applicability to a broader context. This partitioning strategy was designed to optimize the model's effectiveness and ensure its relevance in various scenarios. The split not only enabled thorough training but also provided comprehensive validation using new, unexposed data to assess the model's accuracy. Additionally, Normal were included as a control group in the study.

\begin{table}[h]
\caption{Number of the private dataset used for disease classification training and validation datasets}\label{tab2}%
\begin{tabular}{@{}lllllll@{}}
\toprule
Datasets       & HCM  & CA  & Normal \\
\midrule
Training set   & 169 & 144  & 160 \\
Validation set & 43  & 36   & 40  \\
\botrule
\end{tabular}
\end{table}

\subsection{Evaluation metrics}\label{subsec2}
In our pipeline, we utilize four common metrics to evaluate the performance of our models at each phase: Accuracy, Precision, Recall, and the Micro-F1 Score. These metrics collectively offer a comprehensive assessment of the model’s performance.

Accuracy is a fundamental metric in evaluating classification models, reflecting the model's ability to correctly classify samples. Expressed as a percentage, it shows the ratio of accurately classified samples to the total number of samples. The formula for calculating accuracy is as follows:

\begin{equation}
\label{eq:7}
acc = \frac{TP_i + TN_i}{TP_i + TN_i + FP_i + FN_i}
\end{equation}

Here, $i=[1,2,...N]$ represents different categories. $TP$ (True Positives) and $TN$ (True Negatives) are the number of samples correctly classified as positive and negative, respectively. Conversely, $FP$ (False Positives) and $FN$ (False Negatives) are the number of samples incorrectly classified.

The F1 Score, used to assess the specific classification performance of each category, combines Precision and Recall to provide a more comprehensive measure of effectiveness. This is particularly useful in multi-class scenarios with class imbalances. The Micro-F1 Score, ensuring equal contribution of each sample to the overall evaluation, offers a balanced and precise reflection of the model’s performance across various classes.

The formulas for the F1 Score, Precision, Recall, and the Micro-F1 Score are:

\begin{equation}
\label{eq:8}
F1~score_{i} = \frac{2 \times Precision_{i} \times Recall_{i}}{Precision_{i} + Recall_{i}}
\end{equation}

\begin{equation}
\label{eq:9}
Precision = \frac{\sum TP_i}{\sum TP_i + \sum FP_i}
\end{equation}

\begin{equation}
\label{eq:10}
Recall = \frac{\sum TP_i}{\sum TP_i + \sum FN_i}
\end{equation}

\begin{equation}
\label{eq:11}
Micro{-}F1~score = \frac{2 \times Micro{-}Precision \times Micro{-}Recall}{Micro{-}Precision + Micro{-}Recall}
\end{equation}

By employing these metrics, we achieve a detailed and in-depth understanding of our model's performance in a range of classification scenarios.

In order to make the results of the view classification model more intuitive and clearer, we use the T-Distributed Stochastic Neighbor Embedding (t-SNE) \cite{bib27} method for visualization. T-SNE is a nonlinear dimension reduction technique, which is mainly used to visualize high-dimensional data, understand and verify data or models. T-SNE belongs to manifold learning, which assumes that the data is uniformly sampled from a low-dimensional manifold in a high-dimensional Euclidean space. Manifold learning is to recover the low-dimensional manifold structure from the high-dimensional sampled data, that is, to find the low-dimensional manifold in the high-dimensional space and find the corresponding embedding map, so as to achieve dimensionality reduction or data visualization.

In addition, we use the fivefold cross-validation method. Fivefold cross-validation is a common machine learning model evaluation technique, which is usually used to estimate model performance and generalization ability.

\subsection{Training details}\label{subsec2}
In this study, all models were trained on a shared machine learning platform equipped with 8 A100-SXM4 80GB GPUs, 32 Intel(R) Xeon(R) Silver 4314 CPUs @ 2.40GHz processors, and 8032GB of total memory. The training and in ference processes were conducted utilizing only a single GPU card. We used Python 3.8 (under PyTorch 2.0 with CUDA 11.8) for developing the deep learning models. 

In the view classification phase, training employed an SGD optimizer with a learning rate of 0.0005, a batch size of 24, a weight decay of 0.0005, and a momentum of 0.9. We set 30 training iterations. Data normalization was applied uniformly for faster convergence.

In the feature extraction and disease classification phase, training employed an ADAM optimizer with a learning rate of 0.01, a batch size of 1 and a cross entropy loss weight of 0.1. We set 50 training iterations. Data normalization was applied uniformly for faster convergence.

\section{Results}\label{subsec3}
\subsection{View classification}\label{subsubsec1}
Evaluation metrics were calculated using a test set comprising 1,624 ultrasound images. The view classification model, trained on a dataset of 11,951 ultrasound images, achieved an accuracy of 0.95. Detailed experimental results of this model across different views are presented in Table \ref{tab3}. As shown in Table \ref{tab3}, the model attained F1 scores of 0.98 for A4C, 0.97 for PLAX, 0.84 for PSAX\_MV, 0.86 for PSAX\_MP, and 0.92 for PSAX\_AC.

\begin{figure*}
    \centering
    \includegraphics[width=1\textwidth]{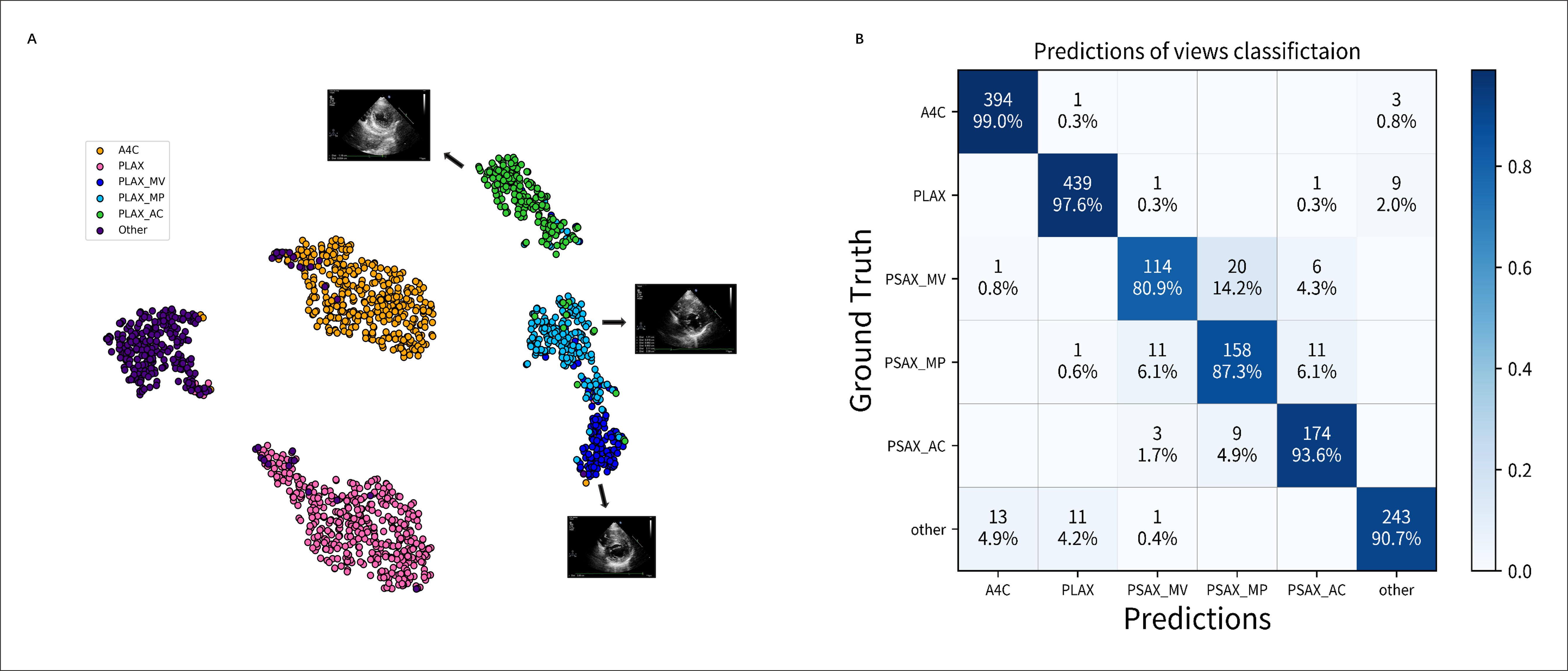}
    \caption{Model of View classification successfully discriminate echocardiographic views. \textbf{A}. T-SNE visualization of view classification depicts the successful grouping of test images corresponding to 5 different echocardiographic views. \textbf{B}. Confusion matrix demonstrating successful and unsuccessful probability of view classifications within the test data set. Numbers along the diagonal represent successful probability of classifications, whereas off-diagonal entries are misclassifications.}
    \label{fig:5}
\end{figure*}

Fig. \ref{fig:5} (A) shows a visual clustering results of the top-level feature extracted by t-SNE, showing the obvious different feature cluster on different view images. The confusion matrix presented in Fig. \ref{fig:5} (B) displays the performance of our view classification model. The model shows high accuracy for A4C view prediction with 394 correct predictions out of 398 (99\%), for PLAX view with 439 correct out of 450 (97.6\%), and  for PSAX\_AC view with 174 correct predictions out of 186 (93.6\%). The model is also able to  identify PSAX\_MV with 114 correct out of 141 (80.9\%), and PSAX\_MP with 158 correct out of 181 (87.3\%). For the 'other' category, it successfully classified 243 out of 268 images (90.7\%). These results indicate a high degree of accuracy across various echocardiographic views.

\begin{table*}[h]
\caption{Experiment results of view classification on the private dataset}\label{tab3}%
\begin{tabular}{@{}l|l|l|l|l|l|l@{}}
\hline
Metrics       & A4C  & PLAX  & PSAX\_MV  & PSAX\_MP & PSAX\_AC & Other\\
\hline
F1 score     & 0.98 & 0.97  & 0.84     & 0.86    & 0.92    & 0.93 \\
\hline
precision    & \multicolumn{6}{c}{0.91}   \\
\hline
recall       & \multicolumn{6}{c}{0.91}  \\
\hline
micro-F1 score & \multicolumn{6}{c}{0.93} \\
\hline
\end{tabular}
\end{table*}

\subsection{Three Type Disease Classification}\label{subsubsec2}
Table \ref{tab4} shows the experimental results using different feature extractions on the five-view private dataset. We observed that ResNet18 and ResNet34 both achieved the highest micro-F1 score of 0.891. ResNet34 had slightly better precision at 0.901 compared to ResNet18's 0.896, and also a marginally higher recall of 0.893, indicating that ResNet34 is the best performing model according to these results. ResNet50, DenseNet121, and EfficientNetB2 demonstrated lower performance, with micro-F1 scores of 0.765, 0.672, and 0.445, respectively. The precision and recall metrics follow a similar trend, with ResNet50 outperforming DenseNet121 and EfficientNetB2, which had the lowest scores across all metrics. 

\begin{table}[h]
\caption{Experiment results of different feature extractor on the private dataset. The best result is in bold}\label{tab4}%
\begin{tabular}{@{}llll@{}}
\toprule
Backbone         & micro-F1 score  & precision  & recall \\
\midrule
ResNet18         & \textbf{0.891}           & 0.896      & 0.889   \\
ResNet34         & \textbf{0.891}           & \textbf{0.901}      & \textbf{0.893}   \\
ResNet50         & 0.765           & 0.771      & 0.766   \\
DenseNet121      & 0.672           & 0.721      & 0.660   \\
EfficientNetB2   & 0.445           & 0.454      & 0.455   \\
\botrule
\end{tabular}
\end{table}

\begin{figure*}
    \centering
    \includegraphics[width=0.8\textwidth]{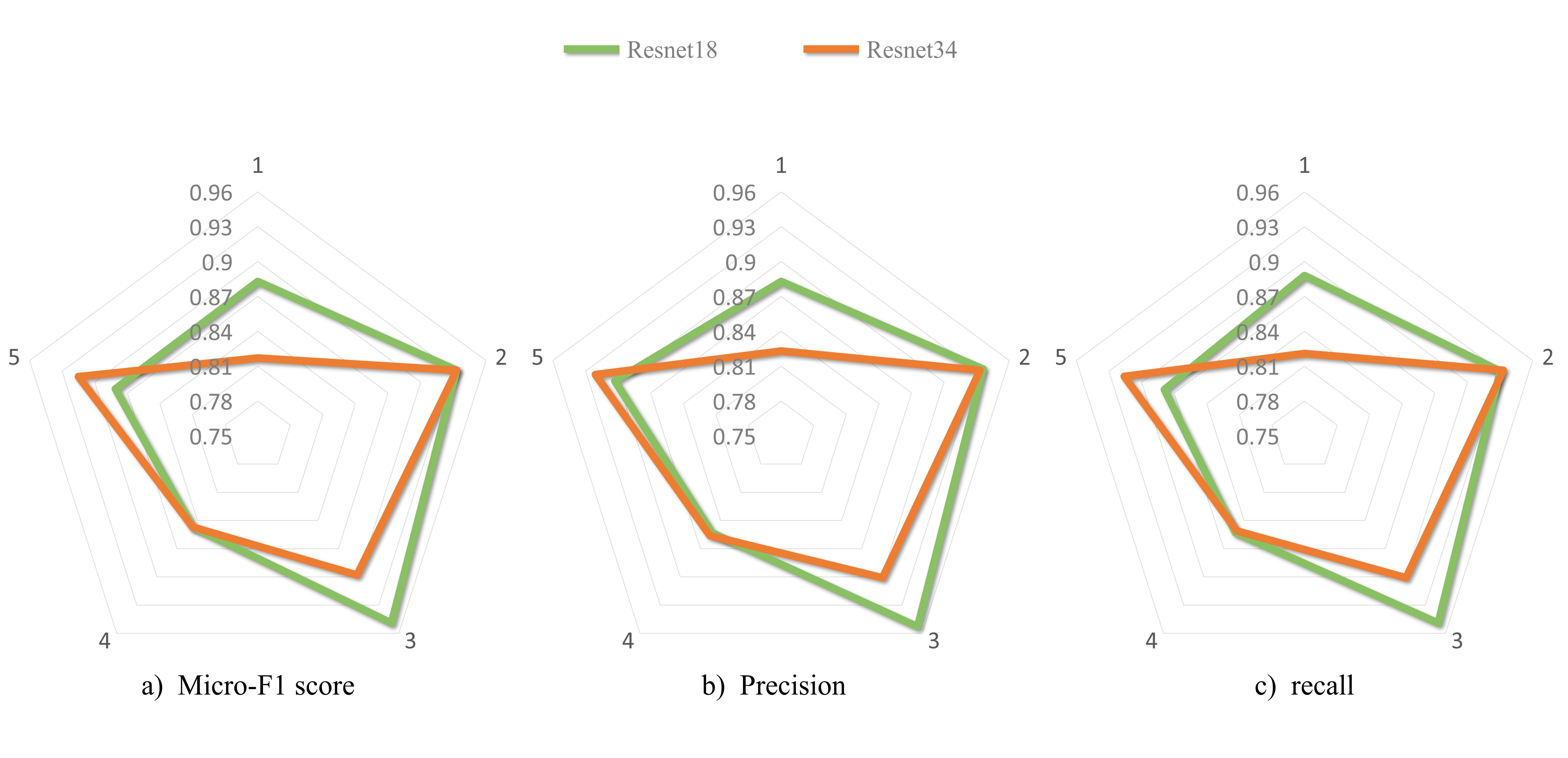}
    \caption{Radar image of fivefold cross-validation of ResNet18 and ResNet34 on the three evaluation metrics. The mean micro-F1 score of ResNet18 and ResNet-34 was 0.904 and 0.882, the mean precision was 0.905 and 0.887, and the mean recall was 0.905 and 0.884, respectively.}
    \label{fig:6}
\end{figure*}

To further compare the efficacy of ResNet18 and ResNet34 as feature extractors, we conducted a fivefold cross-validation focusing on three key metrics. As shown in the radar charts (Fig. \ref{fig:6}), we assessed both models against Micro-F1 Score, Precision, and Recall. The cross-validation revealed that ResNet18 marginally outperformed ResNet34, with a mean Micro-F1 score of 0.904 compared to 0.882, a mean precision of 0.905 against 0.887, and a mean recall of 0.905 versus 0.884 for ResNet34. These results suggest that while both models exhibit commendable performance, ResNet18 demonstrates a slight advantage in all three metrics evaluated.

Table \ref{tab5} shows the results of fivefold cross-validation for various echocardiographic view-based models on our private dataset. The five-view model that we proposed exhibits the best performance with a micro-F1 score of 0.904, precision of 0.905, and recall of 0.905. The results across models indicate that as the number of views increases, the models generally perform better, with the single-view models yielding the lowest scores (A4C: 0.704 for micro-F1 score, 0.722 for precision, and 0.705 for recall; PLAX: 0.738 for micro-F1 score, 0.754 for precision, and 0.741 for recall). The two-view, three-view, and four-view models demonstrate progressive improvements in all three metrics. Notably, the addition of views enhances the model's diagnostic capabilities, culminating in the superior performance of the five-view model. 
\begin{table}[h]
\caption{Experiment results of fivefold cross-validation of different models on the private dataset. The best result is in bold}\label{tab5}%
\begin{tabular}{@{}lllll@{}}
\toprule
Model                              & view                    & micro-F1 score        & precision              & recall \\
\midrule
\multirow{2}{*}{single-view model} & A4c  & 0.704  & 0.722  & 0.705 \\
                                   \cmidrule{2-5}
                                   & PLAX & 0.738  &0.754   &0.741 \\
\midrule
two-view model                  & A4C, PLAX  & 0.821  & 0.840  & 0.823 \\
\midrule
\multirow{3}{*}{three-view model}  & A4C, PLAX, PLAX\_MV & 0.862  & 0.866  & 0.864 \\
                                   \cmidrule{2-5}
                                   & A4C, PLAX, PLAX\_MP & 0.865  &0.884   &0.866 \\
                                   \cmidrule{2-5}
                                   & A4C, PLAX, PLAX\_AC & 0.882  &0.892   &0.882 \\
\midrule
\multirow{3}{*}{four-view model}  & A4C, PLAX, PLAX\_MV, PLAX\_MP  & 0.868  & 0.877   & 0.869 \\
                                   \cmidrule{2-5}
                                   & A4C, PLAX, PLAX\_MV, PLAX\_AC & 0.892  & 0.902   & 0.894 \\
                                   \cmidrule{2-5}
                                   & A4C, PLAX, PLAX\_MP, PLAX\_AC & 0.885  & 0.891   & 0.884 \\
\midrule
\multirow{2}{*}{five-view model (ours)} & A4C, PLAX, PLAX\_MV, PLAX\_MP, & \multirow{2}{*}{\textbf{0.904}}  & \multirow{2}{*}{\textbf{0.905}} & \multirow{2}{*}{\textbf{0.905}}\\ 
                                       &  PLAX\_AC    &  &   & \\

\botrule
\end{tabular}
\end{table}

Table \ref{tab6} presents the experimental results of various disease  classifiers on our private dataset. The linear SVM classifier exhibits the highest scores with a micro-F1 score and recall both at 0.918, and precision also at 0.918. Although the linear network model exhibits a slight decrease in performance compared to the linear SVM, it still maintains robust results with a micro-F1 score of 0.899, precision of 0.906, and recall of 0.900.
Logistic regression, random forest, and XGBoost classifiers exhibit lower performance when compared to linear-SVM and linear network. Logistic regression has a micro-F1 score of 0.848, indicating a moderate predictive capability. Random forest and XGBoost, known for their performance in classification tasks, have micro-F1 scores of 0.811 and 0.843, respectively. These results suggest that ensemble methods like random forest and XGBoost, despite their robustness to overfitting and capacity to model complex interactions, may not always outperform simpler models on every dataset.
\begin{table}[h]
\caption{Experiment results of different disease classifiers on the private dataset.}\label{tab6}%
\begin{tabular}{@{}llll@{}}
\toprule
classifier         & micro-F1 score  & precision  & recall \\
\midrule
linear SVM             & 0.917           & 0.918      & 0.918   \\
logistic regression    & 0.848           & 0.861      & 0.847   \\
random forest          & 0.811           & 0.830      & 0.811   \\
XGBoost                & 0.843           & 0.849      & 0.844   \\
linear network (ours)   & \textbf{0.899}   & \textbf{0.906}  & \textbf{0.900}   \\
\botrule
\end{tabular}
\end{table}

\begin{figure*}
    \centering
    \includegraphics[width=0.5\textwidth]{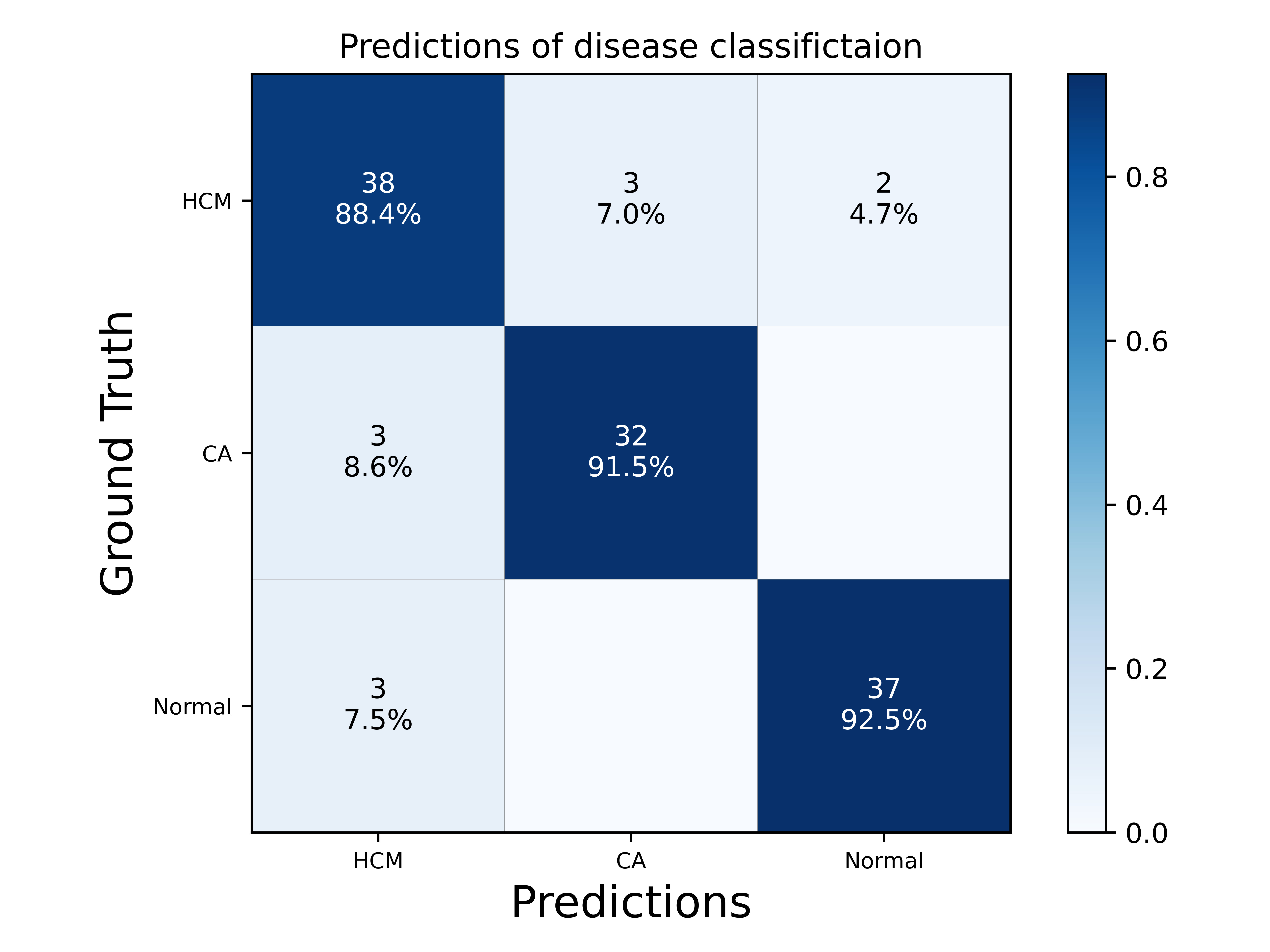}
    \caption{ Confusion matrix of our disease classification model demonstrating successful and unsuccessful probability of disease classification. Numbers along the diagonal represent successful probability of classifications, whereas off-diagonal entries are misclassifications.}
    \label{fig:7}
\end{figure*}

Separate analyses were performed for each category. The confusion matrix presented in Fig. \ref{fig:7} displays the performance of diseases classification model. The model shows high accuracy for HCM prediction with 38 correct predictions out of 43 (88.4\%), for CA with 3 correct out of 35 (91.5\%), and  for Normal with 37 correct predictions out of 40 (92.5\%). And Table \ref{tab7} presents the precision and F1 score of fivefold cross-validation for each category. It is can be seen that the model can effectively differentiate and diagnose HCM and CA with 0.879 of precision for HCM, 0.904 of precision for CA and 0.930 of precision for Normal. 

\begin{table}[h]
\caption{Experiment results of our disease classification model. Using fivefold cross-validation for each category on the private dataset.}\label{tab7}%
\begin{tabular}{@{}llll@{}}
\toprule
                 & precision  & F1 score \\
\midrule
HCM             & 0.879           & 0.877   \\
CA              & 0.904           & 0.908   \\
Normal          & 0.930           & 0.927   \\
\botrule
\end{tabular}
\end{table}

\begin{figure*}
    \centering
    \includegraphics[width=1\textwidth]{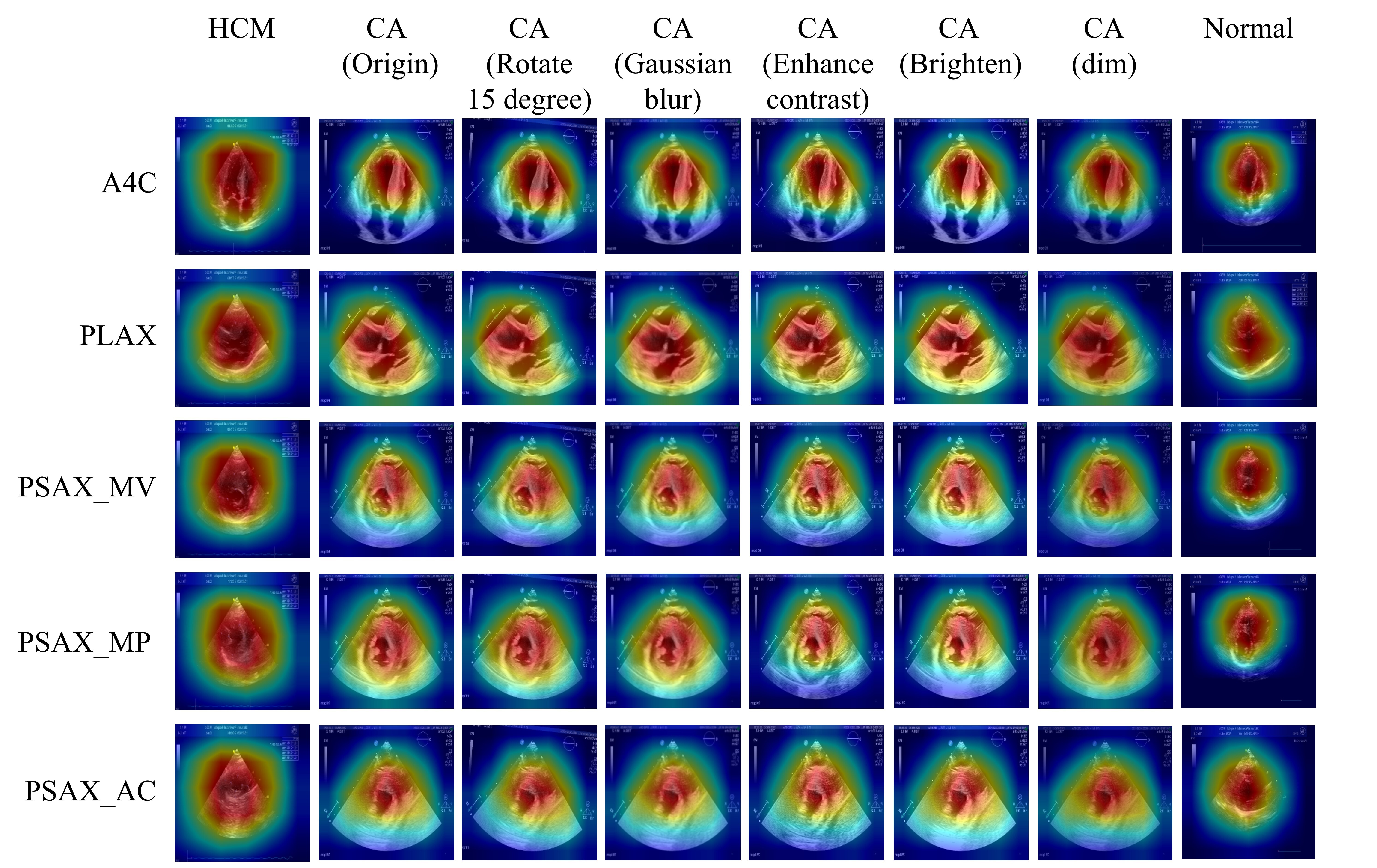}
    \caption{The heat maps of each view generated by Grad-CAM visualizes the ROI that the model focuses on. Color ranking of regions that contributes to the final prediction result of the model: red \textgreater yellow \textgreater blue. The darker the red color, the greater the contribution of the region to the final prediction result of the model, and the higher the attention to this part of the image.}
    \label{fig:8}
\end{figure*}

In order to enhance the interpretability of our model, we employed the Gradient-weighted Class Activation Mapping (Grad-CAM) method \cite{bib30} to visualize heat maps for each view and category, as shown in Fig. \ref{fig:8}. Grad-CAM is a widely used interpretative technique for deep learning models, particularly convolutional neural networks (CNNs), aimed at revealing the regions of interest (ROI) that influence the model's decisions. In our study, Grad-CAM was utilized to highlight the ROIs that the model focuses on in each view of HCM, CA, and Normal categories. Columns 3, 4, 5, and 6 in Fig. \ref{fig:8} depict Grad-CAM visualizations of CA-augmented data. These visualizations demonstrate that, irrespective of whether the data is real or augmented, the model consistently focuses on the heart in all five views, with red areas indicating a particular emphasis on the ventricle and ventricular septum.

It is noteworthy that accurate classification of echocardiographic views is critical for the precise diagnosis of cardiac conditions. As illustrated in Fig. \ref{fig:5}, the views PSAX\_MV and PSAX\_MP are particularly prone to confusion. To highlight the significance of this issue, we conducted an error analysis on these views, chosen for their highest misclassification rate in the confusion matrix. In a simulated experiment involving the deliberate mislabeling of PSAX\_MV as PSAX\_MP in six CA validation set patients, as shown in Table \ref{tab8}, we observed a marked decrease in the probability of correctly diagnosing CA in four patients, with one patient even being misdiagnosed with HCM. This finding emphasizes the importance of accurate view classification in our proposed pipeline, demonstrating that errors in this preliminary step can significantly compromise diagnostic accuracy.

\begin{table*}[h]
\caption{Experiment results of view classification error analysis on the private dataset. The probability of falling is in bold}\label{tab8}%
\begin{tabular}{@{}l|l|l|l|l|l|l@{}}
\hline
Patients of CA   &A1  &A2  & A3  & A4 & A5 & A6\\
\hline
Original probability     & 1.6708  & 1.0504 & 2.3729  & 0.6928 & 1.4779 & -0.6358\\
\hline
Misclassified probability  &2.2449  & \textbf{0.1927}  & 2.4445 & \textbf{-0.5916} & \textbf{0.9952} & \textbf{-1.0293} \\
\hline
Original Result       & CA & CA & CA & CA  & CA & HCM\\
\hline
Misclassified Result  & CA & CA & CA & HCM & CA & HCM\\
\hline
\end{tabular}
\end{table*}

\section{Discussion}\label{sec5}
While several approaches to differentiating HCM and CA have been reported in the literature, our proposed method presents an alternative to these existing techniques. As detailed in Section II, our automatic analysis pipeline is divided into three phases. The initial phase features a view classification neural network module, purpose-built to capture five specific views. This approach contrasts with previous studies, such as one notable example \cite{bib0} that employed a three-tier classification network focusing solely on A4C and PLAX views, and achieved an accuracy of 0.76. Our method, however, utilizes View classification neural networks to automatically acquire A4C, PLAX, PSAX\_MV, PSAX\_MP, and PSAX\_AC views first. The whole pipeline is complemented by a five-input feature fusion neural network architecture and a single-layer neural network classifier, designed for accurate differentiation between HCM and CA.

Our experimental results demonstrate significant advancements. Initially, our classification neural network yielded highly accurate view classification results, with the highest F1 score of 0.98 in A4C and the lowest of 0.84 in PSAX\_MV, alongside a precision and recall of 0.91, and a micro-F1 score of 0.9. This indicates the model's proficiency in obtaining accurate views for subsequent analysis. In the second phase, comparing models trained on single-view to five-view images, the five-view model showed superior performance, underscoring the effectiveness of combining features from multiple views in enhancing disease classification capabilities. The precision of our two-view model is 0.84, which is similar to the model of Yu et al., with 0.76 of accuracy \cite{bib0}. Notably, the five-view model exhibited improvements of 0.078 in micro-F1, 0.066 in precision, and 0.077 in recall compared to the two-view model.

Furthermore, we observed that the linear SVM and linear network classifiers were comparable in disease classification. The choice of a linear network as our preferred model over the linear SVM, despite the latter's marginally higher scores, may be attributed to several factors. Linear networks are neural networks with a single layer that performs a linear transformation. They are scalable and easily integrated with more complex deep learning architectures, allowing for future model enhancements. Moreover, linear networks can benefit from deep learning frameworks' optimizations, GPU acceleration, and they easily fit into a broader deep learning pipeline, unlike SVMs which are traditionally more standalone. Both two classifiers outperformed the other three in terms of micro-F1, precision, and recall. Significantly, as illustrated in Fig. \ref{fig:7}, our three-type classification approach achieved accuracy of 88.4\% for HCM, 91.5\% for CA, and 92.5\% for Normal, highlighting its effectiveness in disease classification.

\begin{figure*}
    \centering
    \includegraphics[width=1\textwidth]{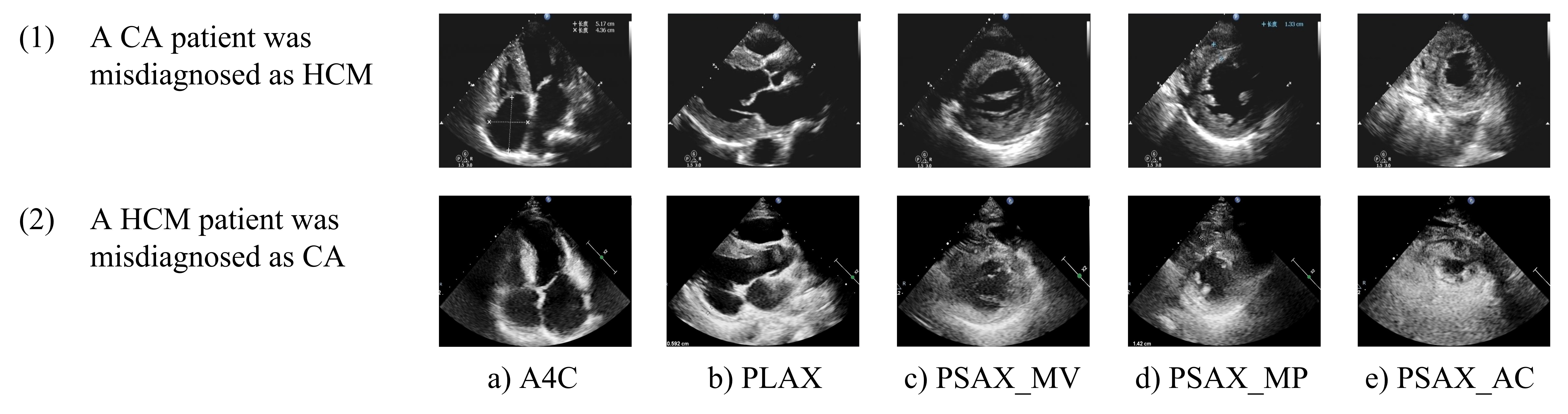}
    \caption{Examples of misdiagnosed patients. (1) is five views of a CA patient misdiagnosed as HCM, while (2) is five views of a HCM patient misdiagnosed as CA.}
    \label{fig:9}
\end{figure*}

In our experimental results, we also identified three cases of CA patients misdiagnosed as having HCM based on echocardiography images. Examination of these images revealed distinct characteristics contributing to these misdiagnoses. For instance, as observed in Fig. \ref{fig:9}, misdiagnosed CA patients exhibited low image gain, poor recognition of myocardial echo features, and thickened myocardium. Conversely, another patient with HCM was misdiagnosed as having CA due to excessively high image gain, poor image quality, and significant echo noise in the myocardium, resulting in an unclear display of the myocardial boundary. Our study yielded an accuracy rate of 0.905 in identifying HCM and CA based on static 5-view echocardiography images, which is competitive compared to diagnoses made by senior sonographers. Referring to the work of Yu et al. \cite{bib0}, our study further selected specific views, including PSAX\_MV and PSAX\_MP and PSAX\_AC, to better reflect the differences between HCM and CA in terms of myocardial echo, thickness, distribution, and pericardial condition. However, as demonstrated by the misdiagnosed cases, slight differences in echocardiography may affect the differentiation between the two conditions.

Multi-modality neural networks have demonstrated superior performance compared to single-modality counterparts in various medical applications \cite{bib35,bib36}. Building on this success, recent research by Wu et al. \cite{bib20} has shown the effectiveness of integrating clinical characteristics, conventional echocardiography, and 2D-STE data in distinguishing between HCM and CA, achieving a remarkable accuracy of 0.98. The fundamental theory underlying this approach involves the calculation of strain and strain rate from estimated displacement obtained through 2D-STE. This technique has been proven effective in distinguishing between HCM and CA \cite{bib31}. It should be noted that, the computational intensity of original 2D-STE poses challenges, particularly in real-time processing. To address this, leveraging CNN-based speckle tracking \cite{bib37} offers avenues for further improvement in the 2D-STE component of Wu et al.'s method. Additionally, to optimize efficiency and accuracy, knowledge distillation strategy could be employed  \cite{bib28,bib29}. This strategy involves utilizing a trained 'teacher' model, such as Wu’s model, to refine the development of our model. By incorporating the advantages of both model inference speed and accuracy, this approach not only enhances the performance of our model but also streamlines the data input process, thus maximizing efficiency and accuracy in medical diagnosis.

\section{Limitations}\label{sec6}
A major limitation of our research is its capacity for generalization, influenced primarily by two factors: the size of the dataset and the integration of multi-center data. The initial challenge is the relatively small scale of the dataset used for preliminary validation. Although effective, a larger dataset covering a wider range of patient demographics is necessary to enhance the robustness and reliability of our model. Expanding the dataset is crucial to minimize overfitting and improve the model's ability to generalize to new cases.

The second challenge involves the complexities of integrating and training with data collected from different centers. Data variability in imaging protocols, devices, and patient characteristics across centers is a significant hurdle for maintaining consistent performance of the model. Future work should concentrate on developing strategies for the effective integration and utilization of multi-center data. Techniques such as federated learning may be key to improving and enhancing the model's performance, ensuring its applicability and accuracy in real-world clinical settings.

\section{Conclusion}\label{sec12}
In this study, we have proposed a deep learning-Driven pipeline for differentiating HCM from CA using 2D multi-View echocardiography. This approach integrates feature information from five distinct cardiac perspectives, facilitating accurate diagnosis and differentiation of these conditions. Experimental results and comparative analysis on a private dataset have substantiated the effectiveness of this feature fusion, encompassing A4C, PLAX, PSAX\_MV, PSAX\_MP, and PSAX\_AC views. This method has demonstrated its capability in the precise discrimination and diagnosis of HCM and CA. Consequently, our research introduces a promising new avenue for diagnosing Cardiovascular Diseases (CVDs), leveraging the comprehensive insights provided by multiple echocardiographic views.

\backmatter



\bmhead{Acknowledgments}

The authors would like to express their gratitude to the High-Performance Computing Center at Southwest Petroleum University for providing GPU support.

\section*{Declarations}

\textbf{Conflict of interest} ~~The authors declare that there are no conflicts of interest to this work.

\bigskip


\bibliography{sn-bibliography}

\end{CJK}

\end{document}